\documentclass[12pt]{iopart}
\usepackage[utf8]{inputenc}
\usepackage{graphicx}

\usepackage{iopams}
\expandafter\let\csname equation*\endcsname\relax
\expandafter\let\csname endequation*\endcsname\relax
\usepackage[comma,authoryear]{natbib}
\usepackage{color,soul}
\usepackage{amsmath}
\usepackage{pdfcomment}

\newcommand{\TCP}{\mathrm{TCP}}

\begin{document}

\title[ ]{The clinical target distribution: a probabilistic alternative to the clinical target volume}

\author{Nadya Shusharina$^1$, David Craft$^1$, Yen-Lin Chen$^2$, Helen Shih$^2$, Thomas Bortfeld$^1$}

\address{
$^1$Division of Radiation Biophysics, Massachusetts General Hospital and Harvard Medical School, Boston, MA 02114, USA

$^2$Department of Radiation Oncology, Massachusetts General Hospital and Harvard Medical School, Boston, MA 02114, USA
}
\ead{nshusharina@mgh.harvard.edu}
\vspace{10pt}
\begin{indented}
\item[]\today
\end{indented}

\begin{abstract}
  Definition of the clinical target volume (CTV) is one of the weakest links in the radiation therapy chain. In particular, inability to account for uncertainties is a severe limitation in the traditional CTV delineation approach. Here, we introduce and test a new concept for tumor target definition, the clinical target distribution (CTD). The CTD is a continuous distribution of the probability of voxels to be tumorous. We describe an approach to incorporate the CTD in treatment plan optimization algorithms, and implement it in a commercial treatment planning system. We test the approach in two synthetic and two clinical cases, a sarcoma and a glioblastoma case. The CTD is straightforward to implement in treatment planning and
  comes with several advantages. It allows one to find the most suitable tradeoff between target coverage and sparing of surrounding healthy organs at the treatment planning stage, without having to modify or redraw a CTV. Owing to the variable probabilities afforded by the CTD, a more flexible and more clinically meaningful sparing of critical structure becomes possible. Finally, the CTD is expected to reduce the inter-user variability of defining the traditional CTV.

\end{abstract}

%
%
%
%
%

\section{Introduction}

The definition of the clinical target volume (CTV) is becoming the weakest link of the radiation therapy chain \citep{austin1995, weiss2003, schlegel2006, njeh2008}. 
The variability in defining the CTV, especially the inter-observer variability, has been documented in numerous publications. A recent systematic review reveals the vast body of literature on this topic published since the year 2000 alone \citep{vinod2016}. The variability/uncertainty in defining the CTV can easily reach one centimeter and more, and it is thus almost one order of magnitude bigger than the millimeter uncertainties in delivering dose to the tumor. 

Recent studies indicate a higher risk of treatment failures in some highly conformal treatments, which could be attributed to inadequately defined CTVs \citep{engels2009}. CTV uncertainties could be particularly detrimental in proton or charged particle therapy, with steeper dose gradients \citep{baumann2016} and without the potentially forgiving ``dose bath'' of conventional radiotherapy. 

There is also a more fundamental conceptual weakness in the current CTV approach: By design, the CTV is a binary concept, which forces the contouring physician to make a yes/no decision as to whether a certain area is tumor target or not. Given the fundamental uncertainties associated with this process, forced binary decisions can lead to some level of arbitrariness in the choice of the CTV, which may explain the variability in defining the CTV. 

There have been other attempts in previous years to deal with CTV uncertainties in a more explicit way. Notably, 20 years ago Waschek et al. proposed a concept where the physicians draw an inner CTV contour which definitely contains tumor, and an outer CTV contour outside of which there is definitely no tumor. 
The vague region between these contours was handled through Fuzzy Logic \citep{waschek1997}. The concept has not been put to practice. This may be in part due to the fact that Fuzzy Logic has not lived up to its promise from the 1980s and 1990s. There have been very few approaches to incorporate CTV delineation uncertainties in treatment plan optimization \citep{balvert2017}, and all of them are based on the standard binary CTV concept. 

In this paper we put forward a continuous probabilistic concept of CTV, which we call the clinical target distribution, CTD. Like the physical dose distribution, the CTD is a three-dimensional discrete distribution. The CTD at a voxel $i$ is the probability $p_i$ that the voxel contains tumor cells.  It generally drops from 100\% near the gross tumor volume (GTV) to 0\% further away from the GTV. 

The CTD can be determined using pathological examinations of microscopic disease extension beyond the GTV  
\citep{vanloon2012, siedschlag2011, akiyama2018analysis}. It can also be guided by CTV contouring studies involving many physicians (by measuring the overlap between them).  It can further be informed by anatomic compartments derived
from diagnostic images, and their known characteristics with respect to tumor spread. 

The main focus of this paper is not how to determine the CTD. The purpose is rather to  
\begin{enumerate}
\item put forward the CTD concept to account for the variable probability of tumor extent beyond the GTV,
\item include the CTD in treatment plan optimization, and
\item demonstrate that the CTD-optimized plans are in some sense superior to conventional CTV-based plans. 
\end{enumerate}

\section{Materials and Methods}
\subsection{Constructing the CTD}
\label{SEC_CREATE_CTD}
The probabilities needed for the CTD distribution can be derived in various ways, as mentioned above, or they can be entered directly by the physician. A realistic scenario is where the physician or a trained imaging algorithm is able to draw and label $H$ shells, from innermost (1) to outermost ($H$) with probabilities $r_h$, where $r_h$ is the probability that there is tumor outside shell $h$, thus  $r_1 > r_2 > \ldots > r_H$. For clarity let us assume the final shell $H$ is drawn such that it is the shell with the smallest probability 
$r_H = 0\%$, i.e. absolute certainty that there are no tumorous voxels outside shell $H$.  An inner shell with, say,  $r_h = 40\%$ means that of 100 patients, 40 would have tumor outside of that shell. 

Now let $s_h = 1-r_h$ be the probability that there is no tumor outside shell $h$. Consider the layer of voxels from shell $H-1$ to shell $H$. Let $q_{H-1}$ be the probability that a voxel in that layer is NOT tumorous, and let the number of voxels in that layer be $N_{H-1}$. Then, under the assumption that the voxels are independent, we have:
\begin{equation}
  s_{H-1} = q_{H-1}^{N_{H-1}}
\end{equation}
\noindent thus $q_{H-1} =  (s_{H-1})^{\frac{1}{N_{H-1}}}$

In general we can solve for all of the $q$ voxel level probabilities by using the relationship:
\begin{equation}
  s_{H-k} = q_{H-k}^{N_{H-k}} s_{H-k+1}
\end{equation}
\noindent which when inverted yields
\begin{equation}
   q_{H-k} = \left( \frac{s_{H-k}}{s_{H-k+1}} \right) ^{\frac{1}{^{N_{H-k}}}}
   \label{q_outside}
\end{equation}
The voxel probabilities of being tumorous $p$ are simply $p=1-q$. 
If the voxels are not independent, these equations have to be modified accordingly. However, the independent voxel assumption appears to be reasonable in absence of further knowledge, and it is the one commonly made in tumor control probability models. 

Note that throughout this paper we consistently distinguish between {\em shells} and  {\em layers}. Shells are infinitely thin surfaces on both sides of layers of voxels. The $h$-th layer of voxels is the region (volume) between shell $h$ and shell $h+1$.
See the example in figure \ref{fig:phantoms}(a). The third layer of voxels (yellow-tinted) is between the third shell $r_3=0.4$ (yellow contour) and the fourth shell $r_4=0.2$ (green contour).

\subsection{Including probabilities in treatment plan optimization algorithms}
The usual way to define the objectives in treatment plan optimization is through the sum of dose-dependent terms (such as piecewise quadratic or piecewise linear terms) $f(d_i)$ in the different voxels: 
\begin{equation}
F = \sum_i f(d_i).
\end{equation} 
We generalize this objective function to the case with voxel probabilities $p_i <1$ by a weighted sum approach:
\begin{equation}
F = \sum_i p_i f(d_i).
\label{F_opt}
\end{equation} 
We will justify this approach below using tumor control probabilities. Note that the only  mathematical difference between the standard CTV and the new CTD methods is the choice of probabilities $p_i$. In the CTV method the $p_i$ equal 1 in the CTV and are 0 elsewhere, whereas in the CTD method the $p_i$ can be arbitrary numbers.

Our general strategy for comparing treatment plans using the new CTD approach with the conventional CTV method is to use the same sets of constraints to keep the critical structures within acceptable dose limits, and the GTV covered by the prescribed dose. The only objective is to keep the dose in the CTD/CTV outside of the GTV as close to the prescription dose as possible. As our voxel-wise objective functions $f(d_i)$ we use quadratic underdose penalties (deviations from the target prescription dose level $d^{\text{ref}}$) of the form: 
\begin{equation} \label{EQ_quadratic_underdose}
f(d_i) \propto  \left[d^{\text{ref}}-d_i \right]_+^2,
\end{equation}
where $[x]_+$ stands for $x$ if $x>0$, and $[x]_+=0$ otherwise.

\subsubsection{Justification using tumor control probabilities}
Here we motivate our pragmatic approach to include the CTD in treatment plan optimization algorithms (equation ({\ref{F_opt}})) through tumor control probability (TCP) models. Let us consider this simple yet widely used TCP model  \citep{webb1993,jin2011}: 
\begin{equation}
\TCP = e^{-n\, e^{-\alpha d}},
\end{equation}
where  $n$ is the number of tumor cells in the finite tumor volume,
$\alpha$ is the sensitivity parameter and $d$ is the radiation dose. Note that we can add a quadratic $(\beta d^2)$ term to the dose, to reflect the linear-quadratic dose effect, but we omit it here to keep the equations simple. 
For example, with $n=10^7$ and $\alpha = 0.35 \, \mathrm{Gy}^{-1}$, we achieve a tumor control level of $\TCP=95$\% with a dose of $d=54.5\, \mathrm{Gy}$. 

Now let us further consider that it is not certain that this volume is actually tumorous. Let the probability of being tumor be $p$, such that the probability of not being tumor is $1-p$. Clearly, if it is not tumor, the TCP equals 1. So then the tumor control probability changes to:
\begin{equation}
\TCP  = (1-p) + p\, e^{-n\, e^{-\alpha d}},
\end{equation}
which is always larger then $1-p$. 

If the overall tumor consists of independent smaller tumor sub-volumes, or ultimately of voxels, we have 
\begin{equation} 
\TCP  = \prod_i \TCP_i = \prod_i \left( (1-p_i) + p_i\, e^{-n_i\, e^{-\alpha d_i}} \right).
\end{equation}

Finally, if every $\TCP_i$ is close to one, which it has to be in order to ensure a reasonably large overall $\TCP$, we can write 
$\TCP_i = 1-\epsilon_i$ with $\epsilon_i$ close to 0. This yields:
\begin{eqnarray} 
\TCP &=  \prod_i(1-\epsilon_i) \\
&= 1 - \sum_i \epsilon_i + \mbox{products and higher-order terms of }\epsilon_i \\
&\approx 1 - \sum_i \epsilon_i ,
\end{eqnarray} 
because the products and higher-order terms are negligible for small $\epsilon_i$. So therefore:
\begin{eqnarray} 
\TCP  &\approx 1 - \sum_i (1- \TCP_i) \\
 &= 1 - \sum_i p_i \left( 1 - e^{-n_i\, e^{-\alpha d_i} } \right). \label{EQ_TCP}
\end{eqnarray}
Note that this is the sum of voxel-level dose dependent terms weighted by voxel probabilities $p_i$, just like equation ({\ref{F_opt}}). Equation  ({\ref{EQ_TCP}}) could be implemented directly as an objective function in optimization algorithms. However, to be more consistent with current clinical practice, we use the quadratic underdose from equation (\ref{EQ_quadratic_underdose}) rather than the double exponential above.

\subsection{Synthetic and clinical cases studied}

\subsubsection{Synthetic cases}
\begin{figure}[ht]
    \centering
    \includegraphics[width=0.8\textwidth]{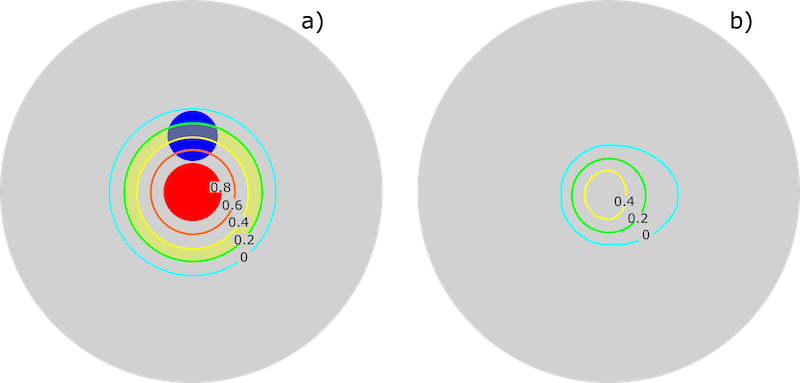}
    \caption{Two geometric phantoms. (a) Five concentric probability shells  distributed equidistantly around a cylindrical target (red) and intersecting OAR (blue). The layer 3 is yellow-tinted in this example. The CTV is the green ($r_4=$0.2) shell. (b) Three anisotropically distributed shells, see text for details. The probability of finding tumorous areas outside of the yellow, green, and cyan shell is 40\%, 20\%, and 0\%, respectively.}
    \label{fig:phantoms}
\end{figure}

To evaluate the feasibility of the CTD approach we created IMRT plans on synthetic CT images comprising a target, organ at risk (OAR), and probability shells.  Two types of geometry were considered, a quasi 2D cylindrical phantom  with isotropic shells and a 3D deformed sphere phantom with anisotropic shells. The isotropic shell phantom was designed with the target as a centrally positioned cylinder of radius 2 cm, with an OAR being a shifted parallel cylinder of radius 1.7 cm  immediately adjacent to the target. The length of both cylinders was 10 cm. Five concentric shells including the surface of the target and separated by 0.9 cm represented probability levels of $r_1=$0.8, $r_2=$0.6, $r_3=$0.4, $r_4=$0.2 and $r_5=$0 (see figure \ref{fig:phantoms}(a)). 

The anisotropic shell phantom was designed as three enclosed 3D shapes. All three shapes consisted of a hemisphere as one part; the second part was one half of an oblate spheroid (inner), a hemisphere (central) and a half of a prolate spheroid (outer). The probability levels of $r_1=$0.4, $r_2=$0.2 and $r_3=$0 were assigned to the inner, central, and outer shapes, respectively (see figure \ref{fig:phantoms}(b)). 

For the symmetric shell phantom, two plans were created, one optimizing the dose coverage of the conventional CTV under the mean dose limiting OAR constraint and another optimizing the coverage of the layers between the shells under the same OAR constraint. For the anisotropic shell phantom, three plans were created, optimizing the dose coverage of the space between shapes under the constraint limiting the integral dose within the whole phantom; three different levels of the integral dose were considered. 

\subsubsection{Clinical cases}
To test how our new approach performs on clinical cases, we created treatment plans for two patients previously treated at our institution, one for a cervical spine (c-spine) chordoma and the other for glioblastoma. There exists a considerable variation among radiation oncologists in defining extended microscopic disease for these tumors. Therefore, we speculate that relaxing a ``solid'' target delineation will help to define the volume requiring prescription dose more consistently across treating physicians.    

\begin{figure}[ht]
    \centering
    \includegraphics[width=0.8\textwidth]{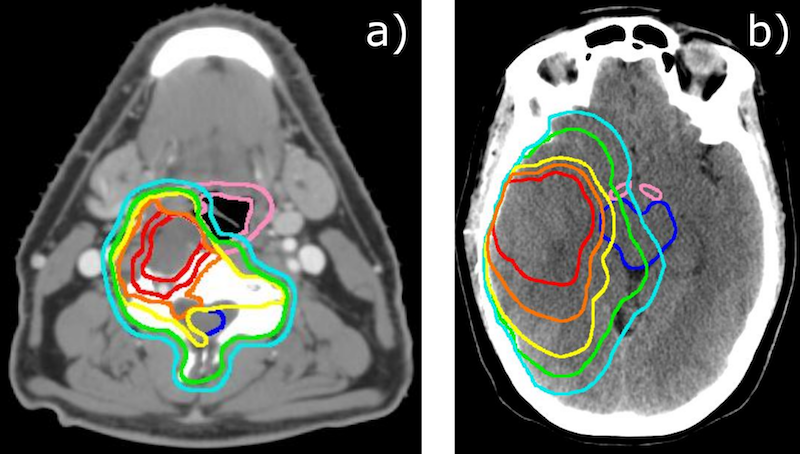}
    \caption{Probability shells delineated on the CT scan of (a) a c-spine chordoma, and (b) a glioblastoma patient. In (a), five shells and the GTV are shown. The red and the orange shells do not encompass the pharynx (pink) or spinal cord (blue),  the next shell (yellow) crosses the pharynx and bends around the spinal cord, and the one matching the CTV (green) and the outermost (cyan) encompass the spinal cord.  In (b), five shells are shown, the surface of the GTV (red), the next two (orange and yellow) which divide the space between the GTV and CTV, the one matching the CTV (green), and the outermost (cyan). The shells are intersecting the brainstem (blue) and optic chiasm (pink).}
    \label{fig:clinical}
\end{figure}

Five probability shells for the c-spine chordoma case were delineated under the guidance of a radiation oncologist and are shown in figure \ref{fig:clinical}(a). The first shell, closest to the GTV, was identified as having probability $r_1$=0.8,
and the voxels in the layer between the GTV and the first shell were weighted with $p_i$=1, the same as the voxels within the GTV (note, voxels given $p_i$=1 are treated as hard constraints by the optimizer: we enforce that they get the prescribed dose). The second shell of $r_2$=0.6 encompassed soft tissues and edges of the vertebral body; it did not cross the surface of the spinal cord or the surface of the pharynx. The third shell with $r_3$=0.4 covered the entire vertebral body and edges of the transverse process; it still did not cross the surface of the spinal cord but intersected the pharynx. The entire spinal cord was encompassed by the next shell with $r_4$=0.2, also enclosed the processes; this shell was identified as CTV for the clinical treatment plan. Finally, the outermost shell with $r_5$=0 was drawn as a 3 mm expansion of the CTV.  Two plans were created: one optimizing the dose coverage of the CTV under the constraints limiting mean dose to the spinal cord and pharynx; and another optimizing the coverage of the layers between shells under the same constraints.   

For the glioblastoma case, the probability shells were drawn based on the physician-defined GTV and CTV. The surface of the GTV was assigned the probability $r_1$=0.8. The space between the GTV and CTV was divided into three layers by drawing two shells with probability levels $r_2$=0.6 and $r_3$=0.4. The CTV was assigned the level $r_4$=0.2. The outermost shell with $r_5$=0 was drawn as a 5 mm expansion of the CTV (see figure \ref{fig:clinical}(b)). Three plans based on probabilistic CTD  were created, optimizing the dose coverage of the space between shells under the constraint limiting the integral dose within the brain; three different levels of the integral dose were considered. In addition, the maximal dose constraints were used to limit the dose to the brainstem and optic chiasm.

\subsection{Implementation in a commercial treatment planning system}
Intensity-modulated radiation therapy (IMRT) plans were generated in the RayStation 5.0 treatment planning system (RaySearch Laboratories, Stockholm, Sweden) using a linear accelerator with 6 MV photons. The dose calculation was performed using the Collapsed Cone algorithm. The optimization objective functions and constraints were used as specified in Table 1.

\begin{table}
\caption{\label{Table1}Plan optimization parameters}
\footnotesize
\begin{tabular}{@{}llll}
\br
Structure & Function & Dose level, Gy & Weight\\
\mr
\it{Isotropic shell phantom}  & & &\\
Target  & Min dose  & 60 & Constraint\\
CTD$^a$     & Max dose  & 66 & Constraint\\
OAR     & Mean dose & 30 & Constraint\\
CTV     & Min dose  & 60 & 1\\
Layer 1 & Min dose  & 60 & 0.58\\
Layer 2 & Min dose  & 60 & 0.20\\
Layer 3 & Min dose  & 60 & 0.09\\
Layer 4 & Min dose  & 60 & 0.05\\
\it{Anisotropic shell phantom}  & & &\\
CTD                 & Max dose  & 63 & Constraint\\
Patient$^b$ - CTD   & Max dose  & 60 & Constraint\\
Patient             & Mean dose & 10, 5, 2.95 & Constraint\\
Layer 0 (inside shell 1)             & Min dose  & 61 & 7.26\\
Layer 1             & Min dose  & 61 & 1.66\\
Layer 2             & Min dose  & 61 & 0.54\\
\it{C-spine chordoma patient} & & &\\
GTV + Layer 0  & Min dose  & 50.4 & Constraint\\
CTD     & Max dose  & 54 & Constraint\\
Patient - CTD   & Max dose  & 52 & Constraint\\
Spinal cord     & Mean dose & 42 & Constraint\\
Pharynx     & Mean dose & 40 & Constraint\\
CTV     & Min dose  & 50.4 & 1\\
Layer 1 & Min dose  & 50.4 & 4.89\\
Layer 2 & Min dose  & 50.4 & 1.19\\
Layer 3 & Min dose  & 50.4 & 0.47\\
Layer 4 & Min dose  & 50.4 & 0.39\\
\it{Glioblastoma patient} & & &\\
GTV  & Min dose  & 62 & Constraint\\
CTD     & Max dose  & 64 & Constraint\\
Patient - CTD   & Max dose  & 62 & Constraint\\
Brainstem     & Max dose & 60 & Constraint\\
Optic chiasm     & Max dose & 54 & Constraint\\
Brain            & Mean dose & 48, 37, 31 & Constraint\\
CTV     & Min dose  & 62 & 1\\
Layer 1 & Min dose  & 62 & 2.21\\
Layer 2 & Min dose  & 62 & 1.78\\
Layer 3 & Min dose  & 62 & 0.65\\
Layer 4 & Min dose  & 62 & 0.32\\
\br
\end{tabular}\\
$^a$the outermost shell; $^b$external contour of a phantom or a patient; \\
Min dose: $f(d_i)=\frac{\Delta v_i}{ d^{\text{ref}}} \left[d^{\text{ref}}-d_i \right]_+^2$, where $d_i$ is the dose in voxel $i$, $d^{\text{ref}}$ is the reference dose level, and $\Delta v_i$ is the relative volume of voxel $i$.
\end{table}
\normalsize

For synthetic cases, all plans utilized 15 equally spaced beams. The beam set for the c-spine case consisted of
9 coplanar beams. For the glioblastoma case, the beam set consisted of 8 beams including 5 coplanar and 3 non coplanar ones. Treatment plan optimization parameters for all cases are listed in Table 1. Optimization cost function included physical dose objectives and constraints. The constraints were chosen to ensure the prescription dose coverage of the GTV, homogeneity of the dose distribution, and to limit the dose to either
the OAR or to the space outside the target. The objectives were chosen to optimize the dose coverage of either the CTV or the probability layers within the CTD. The weight for the only objective function in the CTV-based plan was set to 1.   Optimization objectives for the CTD-based plan were introduced according to equation (\ref{F_opt}), where the weights for voxel layers between probability shells were calculated as probabilities to find  tumor in a voxel belonging to that
layer $p_i=1-q_i$, with $q_i$ defined by equation (\ref{q_outside}). For example, for the four layers of the isotropic phantom, from inner to outer, the number of voxels $N_h$ in each layer for a single slice is $N_1=118$, $N_2=204$, $N_3=313$, and $N_4=447$. The $s$ values are $s_1=0.2$, $s_2=0.4$, $s_3=0.6$, $s_4=0.8$, and $s_5 = 1$. Using equation (3), $q_{H-k}=\left(\frac{s_{H-k}}{s_{H-k+1}}\right)^{\frac{1}{^{N_{H-k}}}}$, we have for $k=1$ (note that $H=5$ in this example): $q_{H-k}=q_4= (s_4/s_5)^{1/n_4} = (.8/1)^{1/447} = .9995$. Other $q$ values are computed similarly yielding the $q$ vector $[0.994, ~0.998, ~0.999, ~0.9995]$. The probability weights are $p=1-q = [0.00585, ~0.0020 , ~0.0009, ~0.0005]$. The values of $p_i$ for each plan were rescaled to be consistent with RayStation's internal hardcoded functions that control plan complexity. These were found through trial and error using guidance from RayStation's solver messages, in particular the phrase ``Optimal solution found'', which indicates good scaling (private communication with RaySearch research staff).
Thus, for example, in Table~1 the $p$ values from the above example are scaled by a factor of 100 to bring them to $[0.58, ~0.2, ~0.09, ~0.05]$. In both the CTV and CTD-based plans the functional form of the objective and sets of constraints were the same as listed in the caption of Table 1.

\begin{figure}[ht]
    \centering
    \includegraphics[width=0.8\textwidth]{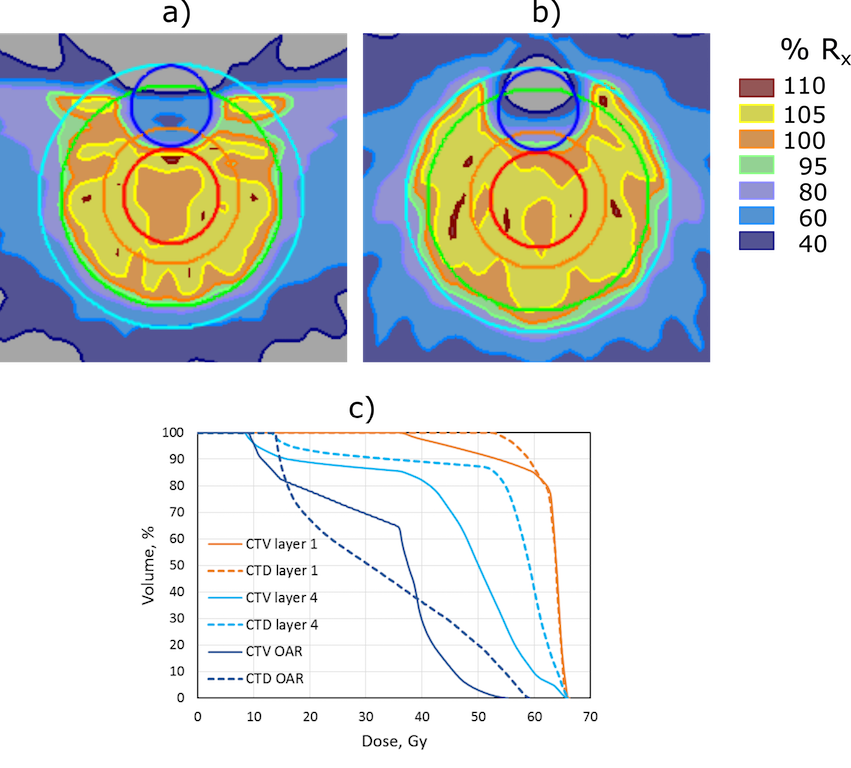}
    \caption{IMRT plans for the geometric phantom with isotropic shells with optimized dose coverage of  (a) CTV (green shell), and (b) CTD with four probability layers. The  innermost, layer 1
      (between the red and orange shells) and outermost, layer 4 (between the green and cyan shells) are shown. The constraint limiting the mean dose to the OAR (blue structure)  was used for both plans. (c) DVHs
      for layer 1 (orange lines), layer 4 (cyan lines), and OAR (blue lines) calculated from the CTV plan (solid lines) and from the CTD plan (dashed lines).}
    \label{fig:synth1}
\end{figure}

\section{Results}

\subsection{Synthetic cases}
Figure \ref{fig:synth1} shows the results for the isotropic shell phantom. The plan that optimizes the coverage of the conventional CTV (figure \ref{fig:synth1}(a)) is compared with the plan that optimizes the coverage of probabilistic CTD with four layers (figure \ref{fig:synth1}(b)). Since the mean dose to the OAR is the same for the two plans, the difference is in the dose coverage of the extended target. As one can see, the dose within  extended target is distributed more uniformly in the case of CTD, with better coverage of the inner and outer layer. The distribution of  dose within OAR is also different, the dose is higher in the region immediately next to the target in the shell-based plan.
The DVHs for the inner- and outer-most layers, and the OAR, calculated from the two plans, are shown in figure \ref{fig:synth1}(c). 

\begin{figure}[ht]
    \centering
    \includegraphics[width=0.8\textwidth]{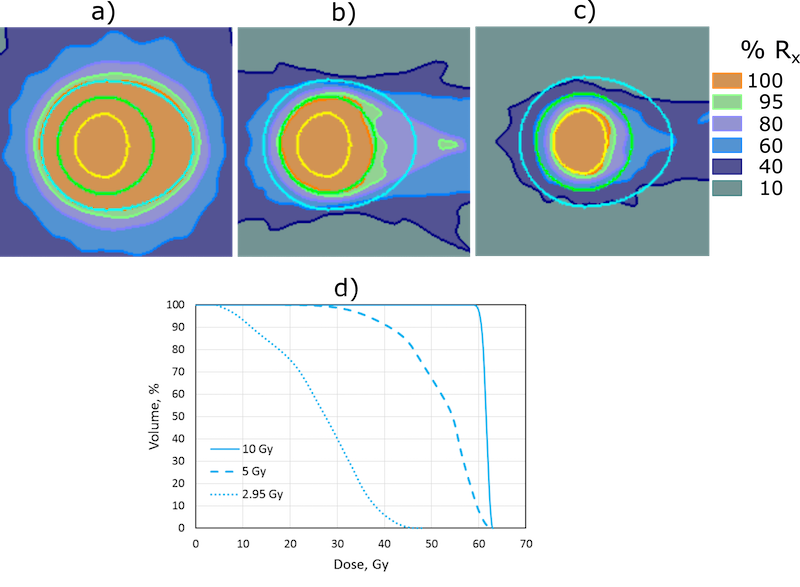}
    \caption{IMRT plans for geometric phantom with anisotropic shells with optimized dose coverage of the CTD under the constraint limiting the integral dose defined as the mean dose within the phantom, $D_{\text{mean}}$. (a) $D_{\text{mean}}=$10 Gy; (b) $D_{\text{mean}}=$ 5 Gy; (c) $D_{\text{mean}}=$ 2.95 Gy; (d) DVHs for the outermost layer (between green and cyan shells) calculated from the three plans with $D_{\text{mean}}=$10 Gy (solid line), $D_{\text{mean}}=$5 Gy (dashed line), and $D_{\text{mean}}=$2.95 Gy (dotted line).}
    \label{fig:synth2}
\end{figure}

In figure \ref{fig:synth2}, the three plans created for the anisotropic shell phantom are compared. Dose coverage of the CTD with three shells was optimized using three different levels of integral dose, defined as the mean dose within the whole phantom, as a constraint. With the integral dose limited to 10 Gy, the prescription dose conforms to the outermost shell of the CTD (figure \ref{fig:synth2}(a)). As the constraint hardens, i.e. the limit for the integral dose decreases to 5 Gy, the dose conforms to the middle shell (figure \ref{fig:synth2}(b)); and when the limit is equal to 2.95 Gy, only the layer within the innermost shell is covered with the prescription dose (figure \ref{fig:synth2}(c)).
The DVHs for the outermost layer, calculated from the three plans, are shown in figure \ref{fig:synth2}(d).   

\subsection{Clinical cases}
\begin{figure}[ht]
    \centering
    \includegraphics[width=0.8\textwidth]{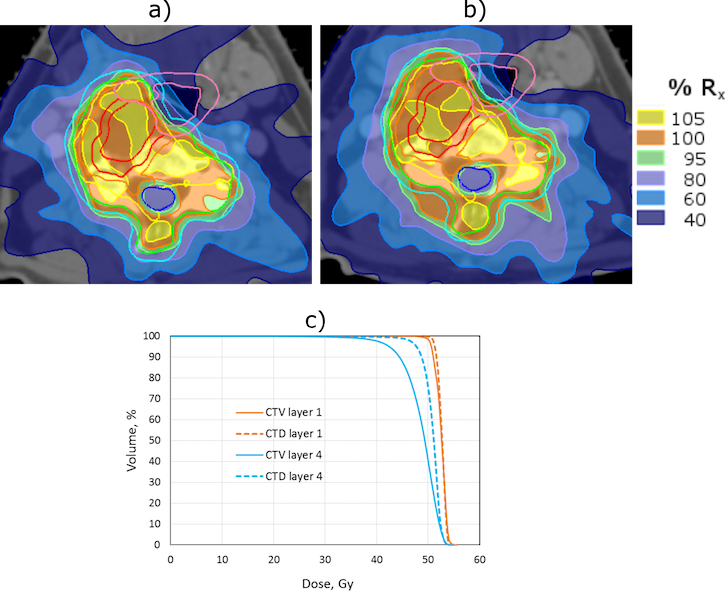}
    \caption{IMRT plans for the c-spine patient with optimized dose coverage of (a) CTV (green shell), and (b) CTD with five probability layers. The plans are optimized to deliver prescription dose to the CTV or CTD  under constraints limiting the mean dose to the spinal cord (blue) and pharynx (pink); (c) DVHs
      for layer 1 (between red and orange shells shown by orange lines and layer 4 (between green and cyan shells) shown by cyan lines calculated from the CTV plan (solid lines) and from the CTD plan (dashed lines).}
    \label{fig:c-spine_plans}
\end{figure}

Plan comparison for the c-spine chordoma patient is presented in figure \ref{fig:c-spine_plans}. Dose coverage of the extended target, CTV or CTD, was optimized under the constraints limiting the mean dose to the spinal cord and the mean dose to the pharynx. The spinal cord was located within the binary CTV delineated by radiation oncologists for the treatment planning. When asked to draw the probability shells, the physician included the spinal cord in the layer between the shells with probabilities $r_3$=0.4 and $r_4$=0.2, with the latter matching the CTV. The pharynx was located immediately next to GTV and intersected binary CTV. The shells were drawn to exclude the pharynx from the first two inner layers and include it into the third layer, so it intersected shells with probabilities $r_3=$0.4, $r_4=$0.2, and $r_5=$0. The proper coverage of the target was challenging since the spinal cord was within the region assigned to receive the prescription dose. Sparing the pharynx was less critical for the coverage requirement but still affected the anterior periphery of the CTV. As shown in figure \ref{fig:c-spine_plans}, both plans achieved the goal of sparing the OARs, although the dose distribution within extended target was different. The probabilistic CTD-based approach was superior to the conventional CTV approach in covering regions marked with higher probabilities of finding tumor cells by shaping the lower dose regions around both spinal cord and pharynx so that the dose fall-off followed the decrease of tumor probability.   

\begin{figure}[ht]
    \centering
    \includegraphics[width=0.8\textwidth]{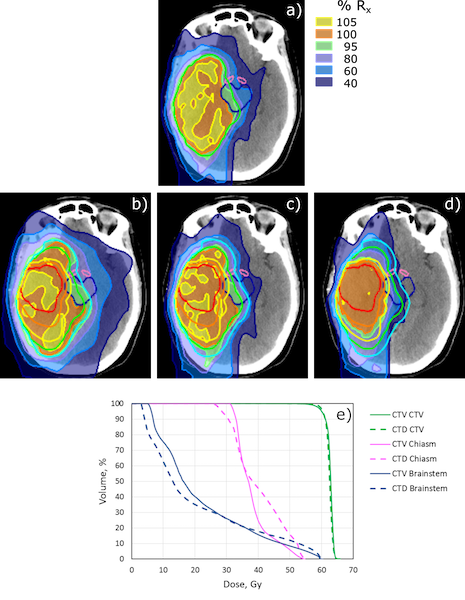}
    \caption{IMRT plans for the glioblastoma patient. The plans are optimized to deliver prescription dose to the CTV (a), and to the CTD with four layers (b-d) under the constraint limiting the integral dose to the brain defined as the mean brain dose, $D_{\text{mean}}$. (a) $D_{\text{mean}}=$ 37 Gy; (b) $D_{\text{mean}}=$ 48 Gy; (c) $D_{\text{mean}}=$ 37 Gy; (d) $D_{mean}=$ 31 Gy.  The additional maximal dose constraint limits the dose to the brainstem (blue) and optic chiasm (pink); (e) DVHs for the CTV (green lines), optic chiasm (pink lines), and brainstem (blue lines) calculated from the CTV plan (solid lines) and from the CTD plan with the same mean brain dose (panel (c)) (dashed lines).}
    \label{fig:gbm_plans}
\end{figure}

IMRT plans created for the glioblastoma patient are shown in figure \ref{fig:gbm_plans}. Panel (a) presents a plan optimized to deliver the prescription dose of 60 Gy to the physician defined CTV, while panels (b)-(d) present plans optimized to deliver the same dose to the CTD with four probability layers. Dose coverage of the CTD was optimized using three different levels of the mean brain dose. The constraints limiting the maximal dose to the brainstem (60 Gy) and optic chiasm  (54 Gy) were used for all plan optimizations. With the mean brain dose of 37 Gy, the prescription dose conforms to the CTV contour in the CTV plan (figure \ref{fig:gbm_plans}(a)) and to the shell that matches the CTV in the CTD plan (figure \ref{fig:gbm_plans}(c)). With the brain $D_{\text{mean}}$=48 Gy the dose conforms to the outermost layer (figure \ref{fig:gbm_plans}(b)).
With the mean brain dose decreased to 31 Gy, the coverage extends only to the shell closest to the GTV (figure \ref{fig:gbm_plans}(d)). The maximal dose constraints for the optic chiasm and brainstem are satisfied for all plans, however, the distributions of the dose within these structures are different for the binary CTV-based optimization and for the layer-based optimization. Comparing DVHs in figure 6(e), it is clear that the layer-based plan optimization results in steeper high dose gradient indicative of higher dose in the proximity of the target.  

\begin{figure}[ht]
    \centering
    \includegraphics[width=0.8\textwidth]{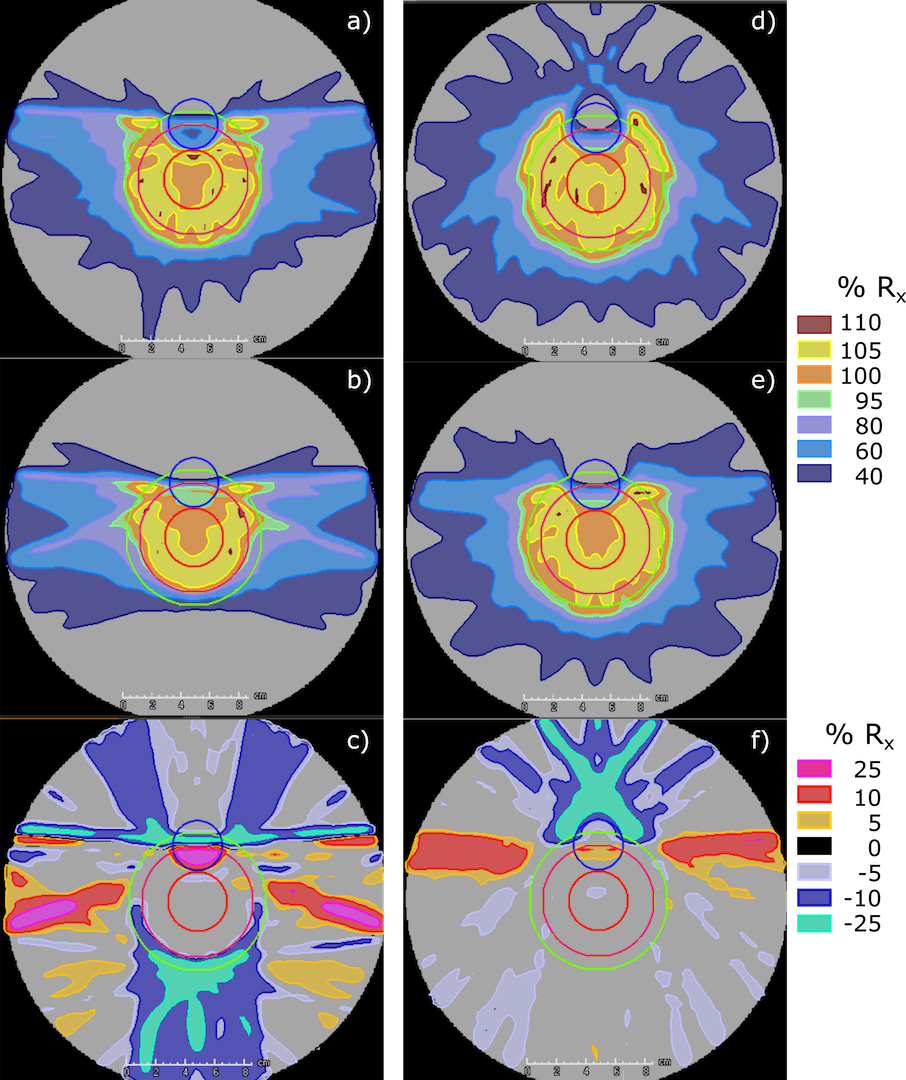}
    \caption{Sensitivity analysis. The left column shows the sensitivity of the CTV-based plan, which results from redefining the CTV contour consistent with how the shell probabilities were altered for the CTD-based sensitivity demonstration. (a) is the original CTV plan, (b) is the revised CTV plan arising from the CTV definition being moved inward one shell, and (c) is the dose difference between these two plans. (d) is the original CTD plan, (e) is the perturbed CTD plan, and (f) is the dose difference map.}
    \label{fig:sensitivity}
\end{figure}

\subsection{Sensitivity analysis}


We demonstrate the sensitivity of the dose distributions to the parameters of the CTD formulation, and at the same time demonstrate that the
CTV approach is in a certain sense more sensitive to these decisions. Consider the target-OAR phantom of figure 1a.
Let $x \in [0,1]$ parametrize the distance from the GTV to the outer shell ($r_5=0$). With that parametrization our nominal formulation
for the decreasing shell probabilities is given by $r(x) = 0.8(1-x)$. Using the rule that the CTV boundary is where $r(x) = 0.2$ we have that
location as 
$x=0.75$. Now we consider a perturbed plan where the system (physician or image-based auto-contouring algorithm) determines that the shell probabilities
decrease faster. In particular we assume $r(x) = 0.8(1-x)^2$. Note this does not alter the shell geometry, just the probabilities. Using this formula and solving for the $x$ location where $r(x) = 0.2$
places the CTV contour at $x=0.5$, i.e. it moves the CTV definition exactly one shell inward. We can now produce two plans, the revised CTV plan and the revised
CTD plan. For the CTD plan, we translate the new $r_h$ probabilities into voxel level probabilities, using equation 3 and scaling by 100, as before. This yields 
$p=[0.85,~ 0.18,~0.055,~ 0.012]$ (in contrast with the nominal values of $[0.58, ~0.2, ~0.09, ~0.05]$). These plans are compared in figure \ref{fig:sensitivity}.
The left column are the CTV-based plans, and the right column are the CTD-based plans. The first row are the original plans, re-displayed here for convenience.
The second row shows the new plans that arise from the perturbed $r_h$ values, and the final row shows the dose differences (new plan minus original plan).
Visual inspection of the plans, as quantified in the dose difference maps, shows that the CTD plans are more similar to each other than the CTV plans, in general and in particular within the target and OAR structures.

\section{Discussion}

We put forward a conceptual framework to deal with uncertainties in defining the clinical target volume by defining it as a continuous probability distribution, CTD. The CTD can be implemented in a straight-forward way in a commercial treatment planning system, as we have shown. 
The concept has a number of advantages. Some of them were demonstrated in this proof-of-principle article, while others are intuitive but have not directly been proven. 

The CTD allows the treatment planner to find the most suitable expansion of the high dose region beyond the visible GTV. Specifically, the tradeoff between covering areas that have a lower probability of being tumorous, and delivering higher doses to the surrounding healthy organs, can be explored interactively by ``navigating'' \citep{muellerMCO} between different options, as shown in figure \ref{fig:gbm_plans}(b)-(d). This is an advantage over conventional planning approaches, which start with the CTV definition and do not permit easy modification of the CTV later during the treatment planning process. Drawing the CTV, the physicians often anticipate the tradeoff between target coverage and sparing of critical organs, which may not reflect the natural physical dose tradeoff. Drawing the target too tight compromises the probability of controlling the tumor, whereas including too generous margins may unnecessarily compromise the function of the surrounding healthy organs. In the CTD case, although the prescription dose still has a binary distribution with a sharp boundary within the CTD, continuous control of the objective function via the CTD will allow the physicians to make a  choice in setting the dose distribution to spare the OARs while covering  the microscopic disease.

The CTD approach can yield better dose distributions. If a certain fraction of the target needs to be excluded from receiving the therapeutic dose level in order to spare neighbouring critical structures, the optimization algorithm can redistribute the dose such that it affects primarily the parts of the CTD with lower probabilities of containing tumor cells, see figure \ref{fig:synth1} and figure \ref{fig:sensitivity}. In other regions where there is no competing OAR nearby, the optimizer will deliver the full prescribed dose even in those voxels with small probabilities of being tumorous. This can be seen in all our examples by looking at the outer CTD layers.


As far as the optimization formulation using probabilities is concerned, we use a first order approximation (neglecting the $\beta D^2$ term of the linear-quadratic formulation) as well as some algebraic simplifications to arrive at the clean voxel probability weighted sum formulation (\ref{F_opt}), but other derivations should be examined in future studies. In other future studies one could examine if a similar voxel-weighted probability approach could be applied in a meaningful way to deal with uncertainties in defining normal tissues.

One expected advantage of the CTD approach is that it will reduce the inter-physician variability of drawing the CTV. We haven't proven this hypothesis but for the following reason it is likely to be true: Consider a case where two physician are unsure whether a certain sub-volume should be included in the clinical target volume or not. Both think the probability that the sub-volume is tumor is of the order of 10\%. Based on that, physician A decides to include it in the CTV, physician B does not. The difference between their CTVs is 100\% in that region, whereas the probabilities, ie, the CTDs, are approximately the same. This fundamental difference--the all-or-nothing nature of CTV-based planning versus the smoothness of the CTD-based approach--is also highlighted in the sensitivity analysis we performed.

We have not addressed the question of how to create the CTD. A few options are mentioned in the introduction. The most realistic short term approach is to have physicians enter an error margin while they draw the CTV. The error margin is then converted to our probability shells as mentioned in section \ref{SEC_CREATE_CTD}. Another option is for them to mark areas of uncertain CTV by directly entering a probability level. 
The ultimate long-term goal is to move away from subjective levels of uncertainty to objective measures of the probability of finding tumor in a voxel. This gold standard can be achieved through pathological studies \citep{vanloon2012, siedschlag2011, akiyama2018analysis}. 

An open question is also how to evaluate the resulting dose distributions, given the uncertain nature of the CTD. In this paper we provided dose volume histograms (DVH) for different shells around the GTV, which is a workable approach. Another option is to create a DVH which plots the {\em expected} volume for each dose bin. 
Here the volume of a voxel with a tumor probability of $p<100\%$  is weighted with that factor $p$. 

We restricted ourselves to the consideration of uncertainties in the CTV definition without considering setup errors, motion, and other sources of spatial uncertainty. Traditionally, those errors are taken into account by adding another margin to the CTV, leading to the ``planning target volume'' (PTV). In the case of the CTD it is not obvious how to add the PTV margin. However, a more advanced recent approach to handle spatial uncertainties is  through robust, probabilistic, and 4D optimization \citep{fredriksson2012, chen2012, trofimov2005, witte2018}, which can be applied in a straight-forward manner to the CTD-based plan. 

To align our approach with common clinical practice, we use a dose based objective function with a fixed prescription dose in our problem formulation, equation (\ref{F_opt}). In particular, we are not using a ``dose painting'' approach \citep{ling2000towards}.  
Like the dependence of the prescription dose on the tumor cell density, the dependence on the tumor probability $p$ is a small logarithmic dependence \citep{webb1993}. This is due to the exponential nature of the cell kill. 
We wish to point out an important difference between variable tumor cell density and variable tumor probability though. Consider a sub-volume of the target where the tumor cell density is only 5\% of the density in the rest of the tumor. This sub-volume requires approximately the same dose as the rest of target. In particular, if we drop the dose to zero in that sub-volume, the TCP drops to zero as well. On the other hand, if the tumor probability of the sub-volume is $5\%$ (i.e., only 5\% of the patients have tumor in that sub-volume), then we can drop the dose to zero and the TCP will only drop by approximately 5\%.

\section{Conclusions}

Moving away from the binary CTV concept towards the more natural and more descriptive CTD concept represents a sizable break from tradition. However, given the potential dosimetric improvements and expected increase in delineation consistency across physicians, we believe that the concept is worth exploring further. Although similar to dose-painting approaches, the CTD concept uses voxel level probabilities--together with treatment plan constraints and objectives -- to determine how each CTD voxel is dosed. It remains to be seen which approach makes more sense in a clinical setting. 

\section{Acknowledgement}
 The project was supported by the Therapy Imaging Program (TIP) funded by the Federal Share of program income earned by Massachusetts General Hospital on C06 CA059267, Proton Therapy Research and Treatment Center.

\newpage

\bibliographystyle{dcu}

\bibliography{CTD.bib}

\end{document}